\documentclass{article}[12pt, a4paper]

\usepackage{amsmath,amsthm, amssymb}

\makeatletter
\renewcommand\section{\@startsection 
{section}{1}{0pt}{-3.5ex plus-1ex minus-.2ex} 
{2.3ex plus.2ex}{\normalfont\large\bfseries}} 
\let\@ldthebibliography\thebibliography
\renewcommand{\thebibliography}[1]{
\centering
        \@ldthebibliography{#1}
}
\makeatother

\newtheoremstyle{theorem}
{10pt} 
{10pt} 
{\sl} 
{\parindent} 
{\bf} 
{. } 
{ } 
{} 
\theoremstyle{theorem}

\newtheorem*{theorem*}{Theorem}
\newtheorem*{lemma*}{Lemma}

\newtheoremstyle{defi}
{10pt} 
{10pt} 
{\rm} 
{\parindent} 
{\bf} 
{. } 
{ } 
{} 
\theoremstyle{defi}

\begin{document}

\title{\bf\large A SCHR\"ODINGER FORMULATION OF BIANCHI I SCALAR FIELD COSMOLOGY}
\author{\normalsize Jennie D'Ambroise\\
\normalsize Department of Mathematics and Statistics\\
\normalsize University of
 Massachusetts\\
\normalsize Amherst, MA 01003, USA\\
\normalsize e-mail:  dambroise@math.umass.edu\\}
\date{}
\maketitle
\thispagestyle{empty}

\noindent{\bf Abstract:}
We show that the Bianchi I Einstein field equations in a perfect fluid scalar field cosmology are equivalent to a linear Schr\"odinger equation.  This is achieved through a special case of the recent FLRW Schr\"odinger-type formulation, and provides an alternate method of obtaining exact solutions of the Bianchi I equations.\\
{\bf AMS Subj. Classification:}  83C05, 83C15

\section{\protect\centering Introduction}
\setcounter{equation}{0}
Recently, a correspondence was established between solutions of Einstein's field
equations in a Friedmann-Lema\^itre-Robertson-Walker (FLRW) universe and solutions of
a particular nonlinear Schr\"odinger-type differential equation.  That is, given a solution of the latter, a solution of the former can be constructed via the prescription given in \cite{JDFW} (and vice versa).
Further motivated by a connection between Bianchi I and FLRW cosmologies seen in a paper by James E. Lidsey \cite{L}, an analogous \underline{linear} Schr\"odinger formulation is demonstrated here for the anisotropic  Bianchi I universe.   The author would like to extend many thanks to Floyd L. Williams for his valued advice on the results presented here.

\section{\protect\centering Einstein Equations}
\setcounter{equation}{0}
Consider the Einstein field equations $T_{ij}=K^2G_{ij}$ for a perfect fluid Bianchi I universe with  scalar field $\phi$, potential $V$ and metric
$ds^2=-dt^2+X(t)^2dx^2+Y(t)^2dy^2+Z(t)^2dz^2$. 
We will consider the case where the energy density and pressure are given solely by a scalar field contribution, i.e. no matter contribution.  That is, $\rho = \dot{\phi}^2/2+V\circ\phi$ and $p =
\dot{\phi}^2/2-V\circ \phi.$  For a vanishing
cosmological constant the equations take the form

\begin{eqnarray}
\frac{\dot{X}\dot{Y}}{XY}+\frac{\dot{X}\dot{Z}}{XZ}+\frac{\dot{Y}\dot{Z}}{YZ} \stackrel{(i)}{=} K^2\rho\\
\frac{\ddot{X}}{X}+\frac{\ddot{Y}}{Y}+\frac{\dot{X}\dot{Y}}{XY} \stackrel{(ii)}{=} -K^2p\notag\\
\frac{\ddot{X}}{X}+\frac{\ddot{Z}}{Z}+\frac{\dot{X}\dot{Z}}{XZ} \stackrel{(iii)}{=} -K^2p\notag\\
\frac{\ddot{Y}}{Y}+\frac{\ddot{Z}}{Z}+\frac{\dot{Y}\dot{Z}}{YZ} \stackrel{(iv)}{=} -K^2p\notag
 \end{eqnarray}
where $K^2=8\pi G$ and $G$ is Newton's constant.

\setlength{\headheight}{1.46cm}

The fluid conservation equation can be derived from these equations and is
\begin{equation*}
\dot{\rho}+\theta(\rho+p)=0\end{equation*}
where
\begin{equation}\theta \equiv \left(\frac{\dot{X}}{X}+\frac{\dot{Y}}{Y}+\frac{\dot{Z}}{Z}\right)\end{equation}
 is the expansion/contraction of volume.  By definitions of  $\rho$ and $p$, this equation reduces to the Klein-Gordon equation
of motion
\begin{equation*}
\ddot{\phi}+\theta\dot{\phi} +V'\circ \phi =0.
\end{equation*}
Note that for
$\gamma=2\dot{\phi}^2/[ \dot{\phi}^2+2(V\circ\phi)]$ one has the equation of state
$p=(\gamma-1)\rho$.

\section{\protect\centering Description of The Correspondence $(X, Y, Z, \phi ,V) \longleftrightarrow u$}
\setcounter{equation}{0}

Similar to the formulation in Lidsey \cite{L}, we first describe how a solution to the Bianchi I equations (i)-(iv) can be used to derive a solution to the FLRW equations.  Since \cite{JDFW} provides the FLRW-Schr\"odinger connection, this will motivate the Schr\"odinger-Bianchi I correspondence. 

We begin by defining the quantities
\begin{equation}\eta_1\equiv\frac{\dot{X}}{X}-\frac{\dot{Y}}{Y}, \quad \eta_2\equiv\frac{\dot{X}}{X}-\frac{\dot{Z}}{Z}, \quad \eta_3\equiv\frac{\dot{Y}}{Y}-\frac{\dot{Z}}{Z}.\end{equation}
Computing $\frac{1}{2}[(i)-(ii)-(iii)-(iv)]$ and using the definitions above, one can verify Raychaudhuri's equation
\begin{equation}\dot{\theta}+2\mu^2-\frac{1}{9}\theta^2+\frac{K^2}{2}(3p+\rho)=0\end{equation}
where $\mu$ is the shear scalar given by $\mu^2\equiv \frac{1}{6}\left(\eta_1^2+\eta_2^2+\eta_3^3\right)+\frac{2}{9}\theta^2$.  Also, rewriting equation (i) using the above notation,
\begin{equation}\frac{5}{9}\theta^2-\mu^2=K^2\rho.\end{equation}

Remarkably, equations (3.2) and (3.3) are a special case of the FLRW field equations (see \cite{JDFW}) with the following substitutions:
\begin{equation}n=6, \quad k=0, \quad a(t)=\left(XYZ\right)^{1/3}, \quad D=\frac{X^2Y^2Z^2}{6K^2}\left(\eta_1^2+\eta_2^2+\eta_3^2\right).\end{equation}
Note that \cite{JDFW} requires $D$ to be constant.  By (i)-(iv) in (2.1) and the following lemma, one can easily show that each of the products $XYZ\eta_i$ for $i=1,2,3$ is a constant function of $t$.

\begin{lemma*}For arbitrary functions $X(t), Y(t),Z(t)>0$ and $f(t)$,
\begin{equation*}\dot{f}+\theta f = 0 \Longleftrightarrow fXYZ \mbox{ is a constant function}\end{equation*}
for $\theta$ as in (2.2) in terms of $X,Y,Z$.
\end{lemma*}
\newpage
\noindent Therefore, given a quintet $\left(X, Y, Z, \phi, V\right)$ solution to (i)-(iv) in (2.1), one can construct a solution to the FLRW field equations by (3.4).  By the converse of Theorem 1 from \cite{JDFW}, one can then construct a solution to the time-independent linear Schr\"odinger equation
\begin{equation}u''(x)+[E-P(x)]u(x)=0.\end{equation}
with constant energy $E$ and potential $P(x)$.  We will see that the Schr\"odinger solutions derived in 
this way will always be such that $E<0$.

In this paper, we show the direct correspondence $(X, Y, Z ,\phi , V)\longleftrightarrow u$
between solutions $(X, Y, Z, \phi, V)$ of (i)-(iv) in (2.1)
and solutions $u$ of (3.5).    This correspondence
provides an alternate method of solving Bianchi I field equations.

With the above notation in place, we can now state the main theorem:
\begin{theorem*}
Let $u(x)$ be a solution of equation (3.5), given $E<0$ and $P(x)$.  Then a
solution $(X, Y, Z, \phi ,V)$ of the Einstein equations (i)-(iv) in (2.1) can be
constructed as follows.  First choose functions $\sigma(t)$,
$\psi(x)$ such that
\begin{equation}
\dot{\sigma}(t) =u(\sigma(t))\,, \quad \psi '(x)^2 =\frac{2}{3K^2}P(x)
\end{equation}
and also constants $c_1, c_2$ such that 
\begin{equation}c_1^2+c_1c_2+c_2^2=-\frac{4E}{3}.\end{equation}
Next define the functions
\begin{equation}R(t)=u(\sigma(t))^{-1/3}\end{equation} and 
\begin{equation}
\alpha(t) =\frac{c_1}{2}\sigma(t), \quad \beta(t) =\frac{c_2}{2}\sigma(t),\quad\gamma(t) =-\alpha(t)-\beta(t).\end{equation}
Then the following quintet solves Einstein's field equations (i)-(iv):
\begin{equation} 
X(t)=R(t)e^{\alpha(t)}, \quad Y(t)=R(t)e^{\beta(t)}, \quad Z(t)=R(t)e^{\gamma(t)},\end{equation}
\begin{equation} \phi(t) = \psi(\sigma(t)), \quad V = \frac{1}{3K^2}\left[ (u')^2 + u^2[E-P]\right]\circ\psi^{-1}.
\end{equation}
Here, in fact, $(X,Y,Z,\phi,V)$ will also satisfy the equations
\begin{eqnarray}
\dot{\phi}^2&=&\frac{2}{3K^2}\left(-\dot{\theta}+\frac{E}{X^2Y^2Z^2}\right)\\
V(\phi(t))&=&\frac{1}{3K^2}\left(\theta^2+\dot{\theta}\right).\end{eqnarray}

 Conversely, let $(X,Y,Z,\phi , V)$ be a solution of equations (i)-(iv) in (2.1), with $\rho$ and $p$ as before.  Similar to (3.6), choose some solution $\sigma(t)$ of the equation
\begin{equation}
\dot{\sigma}(t)=\frac{1}{XYZ}.
\end{equation}
Then equation (3.5) is satisfied for
\begin{eqnarray}
E &=& -\frac{1}{2}X^2Y^2Z^2\left(\eta_1^2+\eta_2^2+\eta_3^2\right), \\
P(x) &=& \frac{3}{2}K^2\left[\dot{\phi}^2X^2Y^2Z^2\right]\circ\sigma^{-1}(x),\notag\\
u(x) &=& \left[\frac{1}{XYZ}\right]\circ\sigma^{-1}(x).\notag
\end{eqnarray}
\end{theorem*}
\noindent Note that in (3.15), $E<0$ and is constant by the same argument stated for $D$ above the lemma.  The theorem therefore provides a concrete correspondence $(X,Y,Z, \phi , V)
\leftrightarrow u$ between solutions $(X,Y,Z, \phi , V)$ of the
field equations (i)-(iv) and solutions $u$ of the linear
Schr\"odinger equation (3.5). 

\underline{Remarks.}
\begin{enumerate}
\item The case $P(x)=0$:\\

 In most examples $P(x)$ is nonzero.  However, if $P(x)=0$ then the theorem must be stated carefully, as $\psi(x)$ is a constant function by  (3.6) and has no inverse.  Therefore the expression for $V$ in (3.11)
 has no meaning.  In this case, we will show that the right-hand side of 
  (3.13) is a constant function that will serve as our new definition for $V$. 
  By (3.8)-(3.10), $u\circ\sigma=1/(XYZ)$.  Differentiating and using (3.6),
  $(u'\circ\sigma)(u\circ\sigma)=-\theta/(XYZ)$.  That is, $u'\circ\sigma=-\theta$.  Differentiating 
  again, $(u''\circ\sigma)(u\circ\sigma)=-\dot{\theta}$.  Using these in (3.5), composed with $\sigma$ and multiplied by $u\circ\sigma$,
  \begin{equation}-\dot{\theta}+\frac{E}{X^2Y^2Z^2}=0.\end{equation}
Therefore (3.12) is still valid, of course, with the left side equal to zero since $\phi$ is constant in this case by (3.11).  Differentiating (3.16),
    \begin{equation}-\ddot{\theta}-\frac{2E\theta}{X^2Y^2Z^2}=0.\end{equation}
  Now, to show that (3.13) is constant, differentiate its right side to get
  \begin{eqnarray*}\frac{d}{dt}\left\{\theta^2+\dot{\theta}\right\}&=&2\theta\dot{\theta}+\ddot{\theta}\\
 &=&2\theta\left(\dot{\theta}-\frac{E}{X^2Y^2Z^2}\right)\mbox{ by (3.17)}\\
 &=&0 \mbox{ by (3.16)}.
 \end{eqnarray*}
  
Therefore in the case $P(x)=0$ when the definition of $V$ in (3.11) no longer has meaning, 
 we define $V(x)=V_0\equiv(\theta^2+\dot{\theta})/(3K^2)$ and $\phi(t)=$ any constant; and we note that equations (3.12) and (3.13) still hold in this special case.
 
 \item Equations (3.12) and (3.13) imply (i)-(iv) in (2.1) under a condition:\\
 
By the comment preceding the lemma above, any solution to the equations (i)-(iv) will have the property that $XYZ\eta_i$  is constant for $\eta_i$ as in (3.1) and $i=1,2,3$.  Suppose we are given \underline{a priori} (positive) functions $X,Y,Z$ with this property.  Differentiating each of $XYZ\eta_i$ and setting equal to zero shows exactly that the left sides of (ii)-(iv) are equal to each other.  Next, any three positive functions can be reparametrized as in (3.10) for $R(t)=\left(XYZ\right)^{1/3}$, $\alpha=\frac{1}{3}\ln{\left(\frac{X^2}{YZ}\right)}$, $\beta=\frac{1}{3}\ln{\left(\frac{Y^2}{XZ}\right)}$ and $\gamma=\frac{1}{3}\ln{\left(\frac{Z^2}{XY}\right)}$.  Using these formulas and the constant quantities $XYZ\eta_i$, one can easily compute that each of $\dot{\alpha},\dot{\beta}$ and $\dot{\gamma}$ are scalar multiples of $1/XYZ$, therefore establishing (3.9) for $\sigma$ as in (3.14).  Finally, \underline{under the condition} that $c_1,c_2$ satisfy (3.7) for the constant $E<0$ given by $X,Y,Z$ as in (3.15),  one can show that equations (3.12) and (3.13) indeed compute $\phi$ and $V$ solving (i)-(iv) in terms of the given $X,Y,Z$.
 
\end{enumerate}

\section{\protect\centering Examples}
As an illustration of the theorem, take the solution  
$u(x)=Ae^{-\sqrt{-E}x}-Be^{\sqrt{-E}x}$ for $A,B >0$ to the equation (3.5) with $E<0$ and $P(x)=0$.
Solving the differential equation (3.6) for $\sigma$ using Mathematica, we obtain
\begin{equation}\sigma(t)=\frac{1}{2\sqrt{-E}}\ln\left[\frac{A}{B}\tanh^2[\sqrt{-ABE}(t-c_0)]\right]\end{equation}
for integration constant $c_0$.
Also by (3.6), $\psi(x)=\psi _0\equiv$ any constant.  Then by (3.8)
\begin{equation}R(t)=u\circ\sigma^{-1/3}=\left(\frac{1}{2\sqrt{AB}}\sinh[2\sqrt{-ABE}(t-c_0)]\right)^{1/3}.\end{equation}
Further let $c_1,c_2$ be {\it any} constants such that (3.7) holds given the constant choice $E$.  We form $X,Y,Z,\phi$ according to (3.9)-(3.11) and obtain
\begin{eqnarray}X&=&
\left(\frac{1}{2\sqrt{AB}}\sinh[2\sqrt{-ABE}(t-c_0)]\right)^{1/3}
\left(\sqrt{\frac{A}{B}}\tanh[\sqrt{-ABE}(t-c_0)]\right)^{c_1/(2\sqrt{-E})}\\
Y&=&
\left(\frac{1}{2\sqrt{AB}}\sinh[2\sqrt{-ABE}(t-c_0)]\right)^{1/3}
\left(\sqrt{\frac{A}{B}}\tanh[\sqrt{-ABE}(t-c_0)]\right)^{c_2/(2\sqrt{-E})}\notag\\
Z&=&
\left(\frac{1}{2\sqrt{AB}}\sinh[2\sqrt{-ABE}(t-c_0)]\right)^{1/3}
\left(\sqrt{\frac{A}{B}}\tanh[\sqrt{-ABE}(t-c_0)]\right)^{-(c_1+c_2)/(2\sqrt{-E})}\notag\end{eqnarray}
for $t>c_0$ and $\phi=\psi_0$.  Since $P=0$, $\psi^{-1}$ does not exist (see Remark 1) and we use (3.13) for the definition of $V$ and obtain the constant $V=V_0\equiv -4ABE/(3K^2)$.  The reader may compare this solution with a similar one in \cite{BJ}.

As another example, we will begin with the same assumptions on $E,P$ and will obtain quite a different Einstein solution.  That is, again let $E<0$ and $P(x)=0$, but take solution $u(x)=Ae^{-\sqrt{-E}x}$ to (3.5) with $A>0$.  Solving the differential equations in (3.6), we obtain $\sigma(t)=\ln\left[A\sqrt{-E}(t-c_0)\right]/\sqrt{-E}$ for $t>c_0$ and $\psi(x)=\psi _0\equiv$ any constant (therefore we will also have $\phi=\psi_0$ by (3.11)).  By (3.8), $R(t)=(\sqrt{-E}(t-c_0))^{1/3}$.  Letting $c_1,c_2$ be any solution to (3.7), finally we compute $X,Y,Z$ to be
\begin{eqnarray}X&=&
A^{c_1/(2\sqrt{-E})}\left(\sqrt{-E}(t-c_0)\right)^{c_1/(2\sqrt{-E})+1/3}\\
Y&=&
A^{c_2/(2\sqrt{-E})}\left(\sqrt{-E}(t-c_0)\right)^{c_2/(2\sqrt{-E})+1/3}\notag\\
Z&=&
A^{-(c_1+c_2)/(2\sqrt{-E})}\left(\sqrt{-E}(t-c_0)\right)^{-(c_1+c_2)/(2\sqrt{-E})+1/3}\notag\end{eqnarray}
by (3.9)-(3.10) for $t>c_0$.
Again, since $P=0$, $\psi^{-1}$ does not exist and we use (3.13) as the definition for $V$ and obtain $V=0$.  That is, this solution is vacuum.  

To further demonstrate the utility of the theorem, we will take a trivial non-physical solution of (i)-(iv), and first use the converse theorem to map to a solution of (3.5).  We will then apply the theorem a second time and map back to a solution of (i)-(iv) and will have produced a physically acceptable example.  Begin by considering the vacuum ($\phi=V=0$) solution 
\begin{equation}X=R_0e^{\alpha_0}, \quad Y=R_0e^{\beta_0}, \quad Z=R_0e^{\gamma_0}\end{equation}
for constants $R_0>0,\alpha_0,\beta_0,\gamma_0$ such that $\alpha_0+\beta_0+\gamma_0=0$.  By (3.1) and (3.15), $E=P=0$ and $u(x)=u_0\equiv(1/R_0^3)$.  Clearly this is a solution to the linear Schr\"odinger equation (3.5).  Note that since $u$ is constant and $\phi$ is zero, we did not need to compute $\sigma$.  Now to map back to a solution of (i)-(iv), we solve (3.6) and use (3.11) so that $\sigma(t)=(1/R_0^3)t$ and $\psi=\phi=$ any constant.  Now by (3.8)-(3.10), 
\begin{equation}X=R_0e^{c_1t/(2R_0^3)},\quad Y=R_0e^{c_2t/(2R_0^3)},\quad Z=R_0e^{-(c_1+c_2)t/(2R_0^3)}.\end{equation}
Again by (3.13), $V=0$.

As a final example, we take $u(x)=(1/x)e^{Ex^2/2}$, $E<0$ and $P(x)=(2/x^2)+E^2x^2$ for $x>0$.  This Schr\"odinger solution was found using the techniques in \cite{Levai}.  Solving (3.6), we obtain $\sigma(t)=\sqrt{-\frac{2}{E}\ln[-E(t-c_0)]}$ for integration constant $c_0$ and 
\begin{equation}\psi(x)=\frac{1}{\sqrt{6}K}\left(\sqrt{2+E^2x^4}+\sqrt{2}\ln\left[\frac{x^2}{2+\sqrt{4+2E^2x^4}}\right]\right).\end{equation}
Graphing this function for a few values of $E$ indicates that the inverse exists, and we will denote it by $\psi^{-1}$.  Calculating $X,Y,Z,\phi,V$ according to the theorem, 
\begin{eqnarray}X&=&
\left((t-c_0)\sqrt{-2E\ln[-E(t-c_0)]}\right)^{1/3}e^{c_1/\sqrt{\ln[-E(t-c_0)]/(-2E)}}\\
Y&=&\left((t-c_0)\sqrt{-2E\ln[-E(t-c_0)]}\right)^{1/3}e^{c_2/\sqrt{\ln[-E(t-c_0)]/(-2E)}}\notag\\
Z&=&\left((t-c_0)\sqrt{-2E\ln[-E(t-c_0)]}\right)^{1/3}e^{-(c_1+c_2)/\sqrt{\ln[-E(t-c_0)]/(-2E)}}\notag\\
V&=&
\frac{1}{3K^2}\left[
-\frac{e^{Ex^2}}{x^4}\left(1+Ex^2\right)
\right]\circ\psi^{-1}(t)\notag\end{eqnarray}
\begin{equation}
\phi=\frac{1}{\sqrt{3}K}
\left(\sqrt{1+2\ln^2[-E(t-c_0)]
}+\ln\left(\frac{-\ln[-E(t-c_0)]
}{E+E\sqrt{1+2\ln^2[-E(t-c_0)]}}\right)\right)\notag
\end{equation}
for $c_1,c_2$ satisfying (3.7) and $t>c_0$.

\end{document}